**A ferromagnetic oxide semiconductor as spin injection electrode in magnetic tunnel junction**

Hidemi Toyosaki, Tomoteru Fukumura[*], Kazunori Ueno, Masaki Nakano, Masashi Kawasaki[†]

*Institute for Materials Research, Tohoku University, Sendai 980-8577, Japan*

A magnetic tunnel junctions composed of room temperature ferromagnetic semiconductor rutile $Ti_{1-x}Co_xO_{2-\delta}$ and ferromagnetic metal $Fe_{0.1}Co_{0.9}$ separated by $AlO_x$ barrier showed positive tunneling magnetoresistance (TMR) with a ratio of ~11 % at 15 K, indicating that $Ti_{1-x}Co_xO_{2-\delta}$ can be used as a spin injection electrode. The TMR decreased with increasing temperature and vanished above 180 K. TMR action at high temperature is likely prohibited by the inelastic tunneling conduction due to the low quality of the amorphous barrier layer and/or the junction interface.

---

[*] E-mail address: fukumura@imr.tohoku.ac.jp
[†] Also at Combinatorial Materials Exploration and Technology (COMET), National Institute for Materials Science, Tsukuba 305-0044 Japan.

Semiconductor spintronics provides rich physics of interplay between charge and spin degrees of freedom of electrons and is expected to contribute to the development of next generation semiconductor technology.[1] Many promising applications of ferromagnetic semiconductors have been already demonstrated for (Ga,Mn)As,[2] but the low Curie temperature ($T_C$) prohibited the operation at room temperature. A possible scenario of increasing $T_C$ of ferromagnetic semiconductors was theoretically proposed to use wide gap semiconductors as host materials.[3] Among them, oxide semiconductors are one of the promising candidates to realize high $T_C$ ferromagnetic semiconductors.[4]

Indeed, both anatase and rutile phases of $Ti_{1-x}Co_xO_{2-\delta}$ were found to be ferromagnetic above room temperature by Matsumoto *et al.* including the part of a present authors.[5,6] We also confirmed that this compound has typical electro-magnetic characteristics of ferromagnetic semiconductors such as anomalous Hall effect and ferromagnetic circular dichroism induced by carrier doping.[7-9] These results strongly support the spin polarized nature of the electrons. Therefore, $Ti_{1-x}Co_xO_{2-\delta}$ is one of the promising materials for room temperature semiconductor spintronics. Next challenge is to demonstrate a heterostructure spintronics device using $Ti_{1-x}Co_xO_{2-\delta}$ as the spin injection electrode, which has been performed for (Ga,Mn)As at low temperature.[10,11]

Here, we report on tunneling magnetoresistance (TMR) action of a magnetic tunnel junction (MTJ) composed of $Ti_{1-x}Co_xO_{2-\delta}$ and ferromagnetic metal $Fe_{0.1}Co_{0.9}$ separated by $AlO_x$ barrier.

A cross sectional view of the MTJ heterostructure, $Fe_{0.1}Co_{0.9}$ / $AlO_x$ / $Ti_{0.95}Co_{0.05}O_{2-\delta}$, is illustrated in Fig. 1(a). Rutile $Ti_{0.95}Co_{0.05}O_{2-\delta}$ layer was grown homoepitaxially on (111) surface of conductive rutile $TiO_2$:Nb (0.05 wt.%) single crystal substrates at 250 °C in $10^{-7}$ Torr of oxygen by laser molecular beam epitaxy using a KrF excimer laser. We routinely observed reflection high energy electron diffraction (RHEED) oscillation during the film growth. The resulting films had atomically flat surfaces with well defined atomic steps of 0.2 nm high. For the device fabrication, $AlO_x$ layer was subsequently deposited by ablating sapphire target at 200 °C in $10^{-4}$ Torr of oxygen. The halo RHEED pattern suggested that $AlO_x$ layer was amorphous phase. After the sample was exposed to air in order to transfer into an electron beam deposition chamber, $Fe_{0.1}Co_{0.9}$ layer was deposited without any intentional surface treatment. The $Fe_{0.1}Co_{0.9}$ layer was then capped with Ag (50 nm) *in-situ* to protect from oxidization and/or process damage. The junction area of 50 × 50 $\mu m^2$ was patterned by photolithography and Ar ion milling. Hard-baked photoresist and Ti/Au layer were used for interlayer insulator and contact electrodes, respectively. For TMR measurements, two-probe technique was employed with a dc-voltage bias source, where the magnetic field was applied along the film plane. The positive bias voltage corresponds to the electron tunneling from $Fe_{0.1}Co_{0.9}$ layer to $Ti_{0.95}Co_{0.05}O_{2-\delta}$ layer.

Figure 1(b) shows magnetization curves normalized by the value at $\mu_0 H = 1$ T for $Fe_{0.1}Co_{0.9}$ and $Ti_{0.95}Co_{0.05}O_{2-\delta}$ thin films at various temperatures. The magnetization curves of $Fe_{0.1}Co_{0.9}$ have square shape with the coercive force as low as 5 mT. Those of $Ti_{0.95}Co_{0.05}O_{2-\delta}$ have significantly higher saturation field, due to the out-of-plane easy magnetization axis, with lager coercive force that becomes smaller with increasing temperature. Figure 1(c) shows magnetic field dependences of the TMR ratio $\Delta R/R(1T)$, where $\Delta R$ is $R(\mu_0 H)-R(1\,T)$. The TMR curve at 15 K shows hysteresis and the hysteresis closes at about $\mu_0 H \sim 0.35$ T. With increasing temperature, the hysteresis reduces and eventually disappears at 60 K coincident with the disappearance of the magnetization hysteresis of $Ti_{0.95}Co_{0.05}O_{2-\delta}$ shown in Fig. 1(b). However, the TMR action itself is sustained up to 180 K as described later. The gradual change of TMR ratio at higher magnetic field reflects the high saturation field of $Ti_{0.95}Co_{0.05}O_{2-\delta}$ layer. The junction resistance and the TMR ratio increase with decreasing temperature, however, the rapid increase of resistance of the TiO$_2$:Nb substrate at low temperature (< 15 K) made it impossible to perform reliable measurement, resulting in the observed maximum TMR ratio of ~11 % at 15 K.

Temperature dependences of the product (*RA*) of junction resistance and junction area in zero field cooling and the maximum TMR ratio ($\Delta R/R(1T))_{max}$ are shown in Fig. 2. *RA* decreases significantly with increasing temperature. The inset of Fig. 2 shows that log(*RA*) is proportional to $T^{-1/4}$ above 200 K, indicating that the variable range hopping conduction dominates the transport at such high temperature regime.[14] Considering that direct elastic tunneling conduction is hardly affected by temperature,[15] the increase of conductance at higher temperature is possibly caused by some defects in AlO$_x$ and/or at the junction interface leading to the inelastic tunneling conduction, as was observed in tunneling through amorphous Si barrier.[16] Indeed, ($\Delta R/R(1T))_{max}$ decreases with increasing temperature and disappears over 180 K, above which the variable range hopping conduction dominates. Since $T_C$ of Ti$_{1-x}$Co$_x$O$_{2-\delta}$ is much higher (> 400 K), the improvement of the junction quality is needed to reduce the inelastic tunneling conduction and to achieve higher operating temperature of the MTJ.

Figure 3(a) shows bias voltage ($V_B$) dependence of junction current (*I*). Nonohmic *I*–$V_B$ curve is asymmetric implying the different barrier height between Fe$_{0.1}$Co$_{0.9}$/AlO$_x$ interface and AlO$_x$/Ti$_{0.95}$Co$_{0.05}$O$_{2-\delta}$ interface. Figure 3(b) shows the $V_B$ dependence of differential conductance (d*I*/d$V_B$). d*I*/d$V_B$–$V_B$ curve is asymmetric that is enhanced at lower temperature, and d*I*/d$V_B$ decreases with decreasing temperature. The significant decrease is seen from 180 K to 150 K (for -0.3 V ≤ $V_B$ ≤ 0.1 V), suggesting a good coincidence with the onset temperature of TMR (Fig. 2).[17] A cusplike structure is observed below 30 K at around 0 V with the voltage width of about 40 mV. This structure has been called as zero bias anomaly and was thought to be caused by excitation of magnon at junction interface.[18,19]

Figure 4 shows $V_B$ dependence of the maximum TMR ratio (($\Delta R/R(1T)$)$_{max}$). ($\Delta R/R(1T)$)$_{max}$ has larger $V_B$ dependence at lower temperature and becomes smaller with increasing $|V_B|$, particularly for the positive $V_B$. ($\Delta R/R(1T)$)$_{max}$ at 150 K becomes nonzero for -0.3 V ≤ $V_B$ ≤ 0.1 V approximately: this $V_B$ range coincides well with that for the significant decrease of d$I$/d$V_B$ at 150 K as seen in Fig. 3(b) as was pointed in earlier studies.[17] ($\Delta R/R(1T)$)$_{max}$ decreases slowly with increasing $|V_B|$ in comparison with a metal-semiconductor MTJ such as MnAs/AlAs/(Ga,Mn)As heterostructure.[20] This result is likely originated from the larger exchange splitting of $Ti_{1-x}Co_xO_{2-\delta}$ than that of (Ga,Mn)As, where the latter exchange splitting is ~100 meV, probably leading to the high spin-polarized current.

In summary, a magnetic tunneling junction using $Ti_{1-x}Co_xO_{2-\delta}$ as the spin injection electrode shows distinct positive TMR with the ratio of ~11 % at 15 K and the TMR action is sustained up to 180 K. Improvement of the junction quality such as reduction of defects in the barrier layer and/or at the interface could lead to the TMR action at higher temperature.

*Note added in proof:* The surface of $TiO_2$:Nb substrates was treated by the procedure of Yamamoto *et al.* (Ref. 21).

Acknowledgements

The authors gratefully acknowledge H. Ohno, F. Matsukura and D. Chiba for discussions. This work was supported by the Japanese Ministry of Education, Culture, Sports, Science and Technology in Japan, Grant-in-Aid for Creative Science Research (14GS0204) and for Young Scientists (A16686019) and for Scientific Research on Priority Areas (16076205), NEDO International Joint Research program (02BR3), and the Sumitomo Foundation. H. T. is supported by Research Fellowship of the Japan Society for the Promotion of Science for Young Scientists.

**Figure captions**

Fig. 1. (a) A cross sectional view of $Fe_{0.1}Co_{0.9}$ / $AlO_x$ / $Ti_{1-x}Co_xO_{2-\delta}$ magnetic tunnel junction (MTJ). (b) Magnetic field dependence of normalized magnetization curves $M(\mu_0H)/M(1T)$ for plane films of $Fe_{0.1}Co_{0.9}$ and $Ti_{0.095}Co_{0.05}O_{2-\delta}$ at $T$ = 15 K, 30 K and 60 K. (c) Magnetic field dependences of TMR ($\Delta R/R(1T)$) for MTJ at $T$ = 15 K, 30 K and 60 K, where $\Delta R = R(\mu_0H) - R(1T)$. The bias voltage ($V_B$) is 0.01 V and the magnetic field is applied along the in-plane. The product of $R(1T)$ value and junction area ($A$) is displayed at each temperature. Abscissa and ordinate for (b) and (c) are shifted for clarity.

Fig. 2. Temperature dependences of $R\,A$ with the bias current ($I_B$) of 0.005 µA in zero-field cooling (ZFC) (solid line) for right axis and the maximum TMR ratio ($\Delta R/R(1T))_{max}$ at $V_B$ = 0.01 V (solid symbol) for left axis. The inset shows $\log(RA)$ vs. $T^{-1/4}$ plot.

Fig. 3. Bias voltage ($V_B$) dependences of (a) current ($I$) and (b) differential conductance ($dI/dV_B$) for the magnetic tunnel junction at various temperatures under zero magnetic field.

Fig. 4. Bias voltage dependences of the maximum TMR ratio ($\Delta R/R(1T))_{max}$ at various temperatures. Abscissa is shifted for clarity.

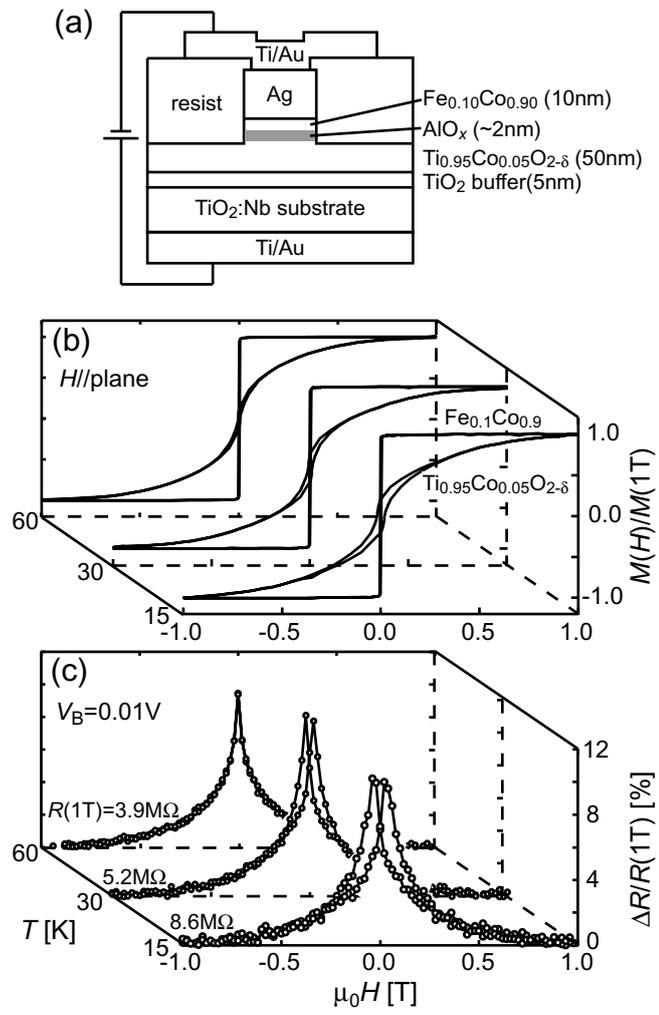

Fig.1 H. Toyosaki *et al.*

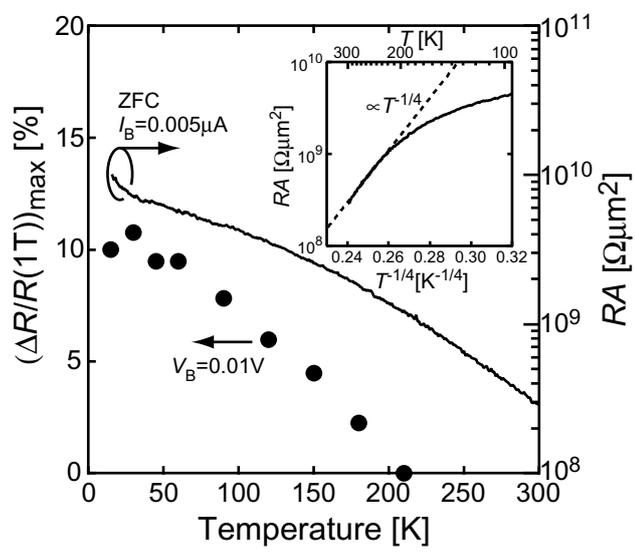

Fig.2 H. Toyosaki *et al.*

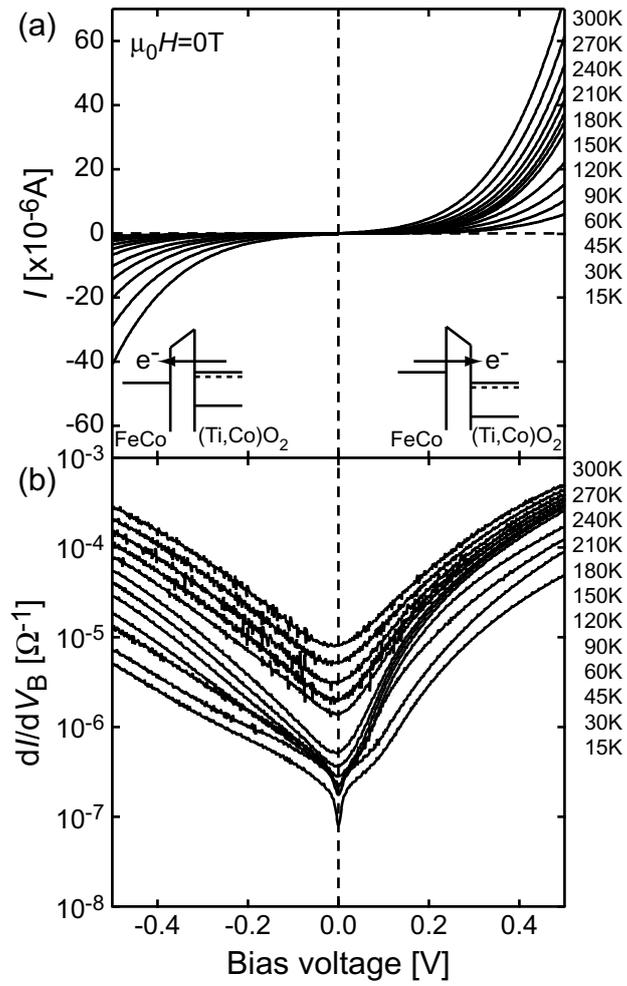

Fig.3 H. Toyosaki *et al.*

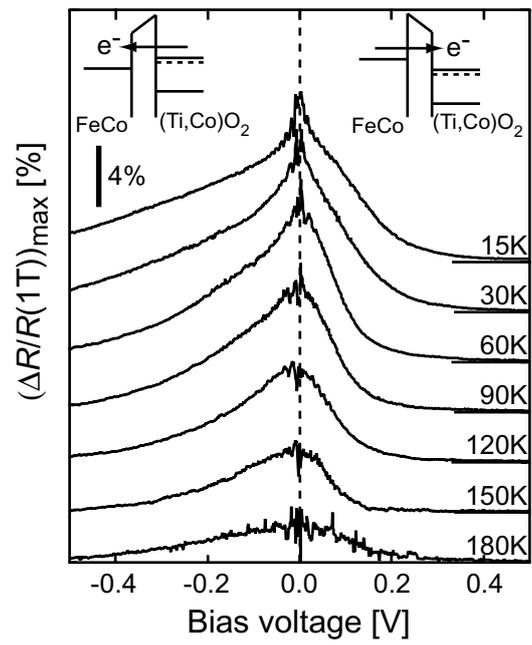

Fig. 4 H. Toyosaki *et al*.